\documentclass[english,aip,apl,amssymb,amsmath,reprint]{revtex4-1}

\usepackage{graphicx}

\begin{document}

\title{Spin and charge thermopower of resonant tunneling diodes}

\author{Javier H. Nicolau}
\affiliation{Institute for Cross-Disciplinary Physics and Complex Systems IFISC (UIB-CSIC), E-07122 Palma de Mallorca, Spain}
\author{David S\'anchez}
\affiliation{Institute for Cross-Disciplinary Physics and Complex Systems IFISC (UIB-CSIC), E-07122 Palma de Mallorca, Spain}


\begin{abstract}
We investigate thermoelectric effects in quantum well systems. Using the scattering approach for coherent
conductors, we calculate the thermocurrent and thermopower both in the spin-degenerate case and in the presence
of giant Zeeman splitting due to magnetic interactions in the quantum well. We find that the thermoelectric current
at linear response is maximal when the well level is aligned with the Fermi energy and is robust against
thermal variations. Furthermore, our results show a spin voltage
generation in response to the applied thermal bias, giving rise to large spin Seebeck effects
tunable with external magnetic fields, quantum well tailoring and background temperature.
\end{abstract}

\pacs{72.20.Pa, 85.75.Mm, 75.50.Pp}

\maketitle

Resonant tunneling diodes are versatile
devices that further enable investigations in fundamental physics.\cite{esaki-RTD-1974} 
Current responses in the GHz regime have been demonstrated
with double-barrier heterojunctions,\cite{sol83,lur85}
which also show bistability \cite{gol87} and super-Poissonian noise \cite{ian98,bla99}
arising from their intrinsic nonlinearities.
These systems are also useful in discussions on coherent versus sequential
scattering processes.\cite{wei87,jon87}
Very recently, tunnel diodes have been used as detectors
of hypersonic wave packets.\cite{you12}

Temperature effects dramatically alter the behavior of
resonant tunneling devices. The peak-to-valley
ratio in the current--voltage characteristics quickly
decreases as temperature increases.\cite{sol83,bon85}
Hole transport has been shown to be quite sensitive
to thermal variations.\cite{men85}
Furthermore, temperature can tune the transition
from static domains to self-sustained oscillations
in multiple-quantum-well structures.\cite{wan99,san01}
However, these works consider a fixed background temperature
common to both the sample and the leads.
More interesting is the generation of thermovoltages
in response to temperature \textit{gradients}
applied to the attached reservoirs (the Seebeck effect).\cite{book}
Recent works suggest that significant improvements
of the heat-to-energy conversion efficiency
can be obtained with low dimensional
systems in general \cite{Effect-QW-on-ZT,BestTE,influence-dimension-ZT}
and with quantum-well tunnel devices in particular.\cite{Nature-first-ZT-super,sot13}
Therefore, it is important to investigate the thermopower properties
of a resonant-tunneling double-barrier system, which have been
little explored up to date.

Our findings reveal a thermocurrent peak when the quantum
well level is aligned with the leads' Fermi energy. This is in stark
contrast with the typical quantum dot behavior, for which
the thermoelectric conductance vanishes at the electron-hole
symmetry point.\cite{sta93,dzu97,sofia} We attribute this difference
to the crucial contribution from the transversal energy
degrees of freedom in tunnel diodes. Furthermore, we
consider giant Zeeman effects arising from diluted
magnetic impurities \cite{slo1,slo2} and find significant values
of the spin bias voltage created from temperature differences
(the spin Seebeck effect)\cite{Uchida,bau10} and highly tunable
with the well level position or the base temperature.

\begin{figure}
	\includegraphics[scale=0.40]{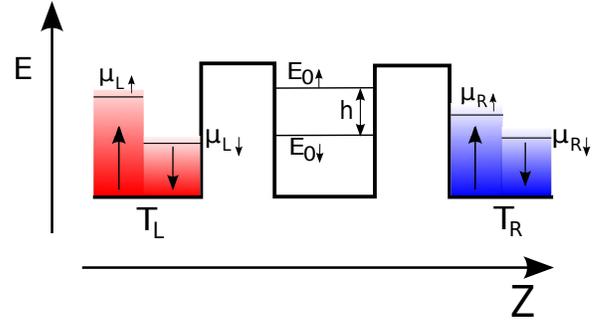}
 	\caption{(Color online) Sketch of a double-barrier tunnel device.  Transport occurs along the $z$ direction.
 	$E_0$ is the spin-split level in the quantum well coupled
 	to left ($L$) and right ($R$) reservoirs with spin-dependent chemical potentials $\mu$
 	and two different temperatures, 	$T_L$ and $T_R$.}
 	\label{fig: RTD}
\end{figure}

Consider a semiconductor heterostructure with two potential barriers
and a quantum well sandwiched between them,
as sketched in Fig.~\ref{fig: RTD}. Quite generally, the energy levels
are spin split due to an external magnetic field to be specified below.
The mean electrochemical potential at lead $\alpha=L,R$ is given
by $\mu_{\alpha}=(\mu_{\alpha \uparrow}+\mu_{\alpha \downarrow})/2$
and the bias voltage between the two contacts is $V=(\mu_L-\mu_R)/e$.
We denote with $E_F$ the common Fermi energy. In the quantum well,
we consider a single subband $E_\sigma=E_\perp+E_0+\sigma h/2$ only
because in tunnel diodes with narrow wells the level spacing is quite large
(e.g., 550--750~meV in Ref.~\onlinecite{bon85}).
Here, $E_\perp$ is the energy associated to the lateral modes perpendicular
to the current direction, $E_0$ is the level position measured from the well bottom,
$\sigma=+$($-$) for spins $\uparrow$($\downarrow$) with the spin quantization
axis taken along the magnetic field direction,
and $h$ is the giant Zeeman splitting
of the order of 10~meV for small fields around 1~T.\cite{slo1}
This splitting arises from the combined effect of an in-plane magnetic field
and diluted magnetic impurities present in the quantum well.
We remark that for the same field strengths, the spin splittings in the nonmagnetic
leads are negligible. Therefore, possible spin biases will emerge from the application
of thermal gradients, as shown below.

Within the scattering approach, the electronic current per spin is given by
\begin{equation}\label{eq: landuer formula}
I_\sigma=\frac{e}{h}\int {\cal T}_\sigma(E)\left[f_L(E)-f_R(E)\right]dE,
\end{equation}
where ${\cal T}_\sigma (E)$ is the transmission function for a carrier with spin $\sigma$
and energy $E=E_\perp+E_z$,
and $f_{\alpha}\left(E\right)=1/1+e^{\left(E-\mu_{\alpha\sigma}\right)/k_B T_{\alpha}}$
is the Fermi-Dirac function for left and right contacts with
temperature $T_{\alpha}$. Neglecting interfacial roughness effects,
the lateral momentum is conserved during tunneling and the transmission thus
depends only on the energy parallel to the current direction, $E_z$.
We integrate Eq.~\eqref{eq: landuer formula} over the transversal modes and find
\begin{align}
\label{eq: current up nolin}
I_{\sigma}&={\cal C} \int \!  dE_z \   {\cal T}_{\sigma}\left(E_z\right)
\left[T_L \ln \left(1+e^{\left(\mu_{L \sigma} - E_z \right)/k_B T_L}\right)\right.\nonumber \\
& -  \left. T_R \ln \left(1+e^{\left(\mu_{R \sigma} - E_z \right)/k_B T_R}\right)\right]\,,
\end{align}
where ${\cal C}=em^* Ak_B/4\pi^2 \hbar^3$ with $m^*$ 
the carrier effective mass and $A$ the device cross sectional area.

The total current is $I=I_++I_-$. Consider for the moment the spin-degenerate case ($h=0$).
We focus on the linear response regime because Seebeck nonlinearities appear only for transmission
line shapes which depend strongly on energy.\cite{nonlinear-TE-David,jacquod,whitney,hershfield} Hence,
the current can be linearized as $I=GV+L\Delta T$,
where $\Delta T=T_L-T_R$ is the temperature difference between the two contacts
and the transport coefficients are
\begin{align}\label{eq: G general}
G&=\frac{2 e{\cal C}}{k_B} \int \!  dE_z \   {\cal T}\left(E_z\right) f\left(E_z\right) \,,\\
L&= 2 {\cal C}\int \!  dE_z   {\cal T}\left(E_z\right) \nonumber\\
&\times\left[\frac{E_z-E_F}{k_B T} \ f\left(E_z\right) +  \ln \left(1+e^{\left(E_F-E_z \right)/k_B T}\right)\right]\,,
\label{eq: L general}
\end{align}
where $T=(T_L+T_R)/2$ is the base temperature.
Importantly, the thermopower $S=-(V/\Delta T)_{I=0}=L/G$ is independent of $m^*$ and $A$.

Equations~\eqref{eq: G general} and~\eqref{eq: L general} are completely general for elastic
transport and arbitrary transmission functions. For definiteness, we consider the Breit-Wigner
approximation and model the transmission as ${\cal T}(E_z)=\Gamma_L \Gamma_R/[(E_z-E_0)^2+(\Gamma_L +\Gamma_R)^2]$,
where $\Gamma_\alpha$ is the level broadening due to coupling to lead $\alpha=L,R$.
Without loss of generality, we consider symmetric barriers, $\Gamma_L=\Gamma_R=\Gamma/2$.

\begin{figure}
	\includegraphics[scale=0.40]{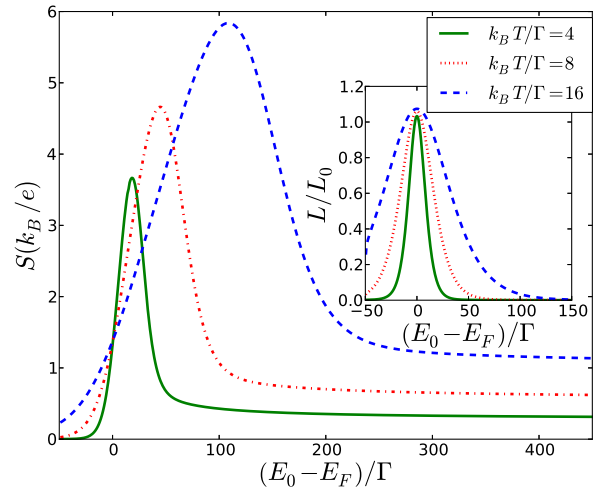}
 	\caption{(Color online) Seebeck coefficient $S$ and thermoelectric conductance $L$ (inset) for a resonant tunneling diode as a function of the quantum well energy level $E_0$. $L$ is normalized to $L_0=em^* A k_B \Gamma/2\pi^2 \hbar^3 $.
 	A maximum is found at $E_0=E_F$ independently of $k_B T/\Gamma$. Both $S$ and $L$ peaks
 	have a width proportional to $k_B T/\Gamma$. }
 	\label{fig: seebeck}
\end{figure}

Figure \ref{fig: seebeck} shows the thermoelectric conductance $L$ (inset) as a function of the relative
position of the energy level in the quantum well. Strikingly, $L$ reaches a maximum when $E_0$
is aligned with $E_F$ irrespectively of the temperature $T$. The background temperature
enhances the peak broadening. This sharply differs with a double-barrier tunnel system
in effectively one dimension (a quantum dot), for which the thermoelectric conductance
vanishes at the particle-hole symmetry point and reaches a maximum (minimum)
when the $E_0-E_F$ is of the order of $\Gamma$ ($-\Gamma$).\cite{sta93,dzu97,sofia}
In our case, $L$ is always positive, implying that the current direction is completely
determined by the sign of the temperature difference. In other words,
electrons are always transported from the hot to the cold side at $V=0$.
In contrast, molecular junctions and quantum dots can exhibit
flow against the thermal gradients when electrons below $E_F$ (holes) are dominant.\cite{molecular}
This is not possible for a quasi-three dimensional tunnel diode because energy is distributed
not only along the transport direction but also among lateral momenta in the perpendicular plane.
 
Accordingly, the thermopower $S=L/G$ attains positive values only, as shown in the main
panel of Fig.~\ref{fig: seebeck}. Therefore, the generated thermovoltage always counteracts the
applied temperature difference and its sign cannot be changed with tuning the well below
or above the Fermi energy. We find that $S$ reaches values as large as $0.5$~mV/K
for quantum levels far beyond $E_F$. The Seebeck coefficient is quite sensitive
to modifications of the base temperature: the maximum position shifts to higher energies
and the peak quickly broadens. When a typical value is used ($\Gamma=1$~meV),\cite{slo2}
$T$ changes in Fig.~\ref{fig: seebeck} from 46~K to 186~K. Of course, for these temperature values
one would expect contributions from inelastic scattering due, e.g., to interaction with phonons
which are neglected in our model. Nevertheless, our results suggest a significant change in $S$
that should be observable at moderate temperatures. In fact, we find that the thermopower
maximum increases proportionally to $ T^{0.33}$ for $\Gamma=1$~meV and its position as a function of $E_0$ is approximately given by $E_F$ when $T\lesssim O \! \left(10^1\right)$~K and by
$7.5k_B T$ when $T\gtrsim O \! \left(10^2\right)$~K.

We now turn to spintronic effects. Recently,
a thermal gradient applied to a metallic ferromagnet was shown to generate
a spin voltage (which produces a spin current) detected from a spin Hall effect
signal.\cite{Uchida,bau10} This discovery has motivated the study of related
phenomena in phase-coherent systems.\cite{swi09,lu10,spin-seebeck-QD} Here, we consider
a magnetically doped quantum well which exhibits a large spin splitting
(denoted with $h$) in the presence of low magnetic fields.\cite{slo1}
In the linear response regime, the total current is $I=\left(L_{\uparrow}+L_{\downarrow}\right) \Delta T + \left(G_{\uparrow}+G_{\downarrow}\right) V + \left(G_{\uparrow}-G_{\downarrow}\right) V_S/2$
while the electronic spin current is $I_S=\left(L_{\uparrow}+L_{\downarrow}\right) \Delta T + \left(G_{\uparrow}+G_{\downarrow}\right) V + \left(G_{\uparrow}-G_{\downarrow}\right) V_S/2$,
where $V_S=\left[\left(\mu_{L \uparrow }-\mu_{L \downarrow }\right)-\left(\mu_{R \uparrow }-\mu_{R \downarrow }\right)\right]/e$
represents the spin voltage. The spin-dependent responses can be obtained
from Eqs.~\eqref{eq: G general} and~\eqref{eq: L general} substituting
$E_0$ with $E_0+\sigma h/2$ in the transmission Lorentzian function.
As a consequence, the Seebeck coefficient becomes split as $h$ is of the order
of the thermopower peak (not shown here).

More interesting is the spin thermopower $S_S$ since it
determines the possibility to create a spin
voltage from a temperature bias only:
\begin{equation}\label{eq: def spin seebeck}
S_S=-\left.\frac{V_S}{\Delta T}\right|_{I=I_S=0}
=\left(\frac{L_{\uparrow}}{G_{\uparrow}}-\frac{L_{\downarrow}}{G_{\downarrow}}\right)\,.
\end{equation}
We note in passing that an electric (charge) voltage
$V=-(L_\uparrow/G_\uparrow+L_\downarrow/G_\downarrow)\Delta T/2$ is to be applied
in order the conditions $I=0$ and $I_S=0$ to be fulfilled. An experimentally more accessible
alternative considers $I=0$ and $V=0$ but it then results in a nonzero spin current.
\begin{figure}
	\includegraphics[scale=0.50]{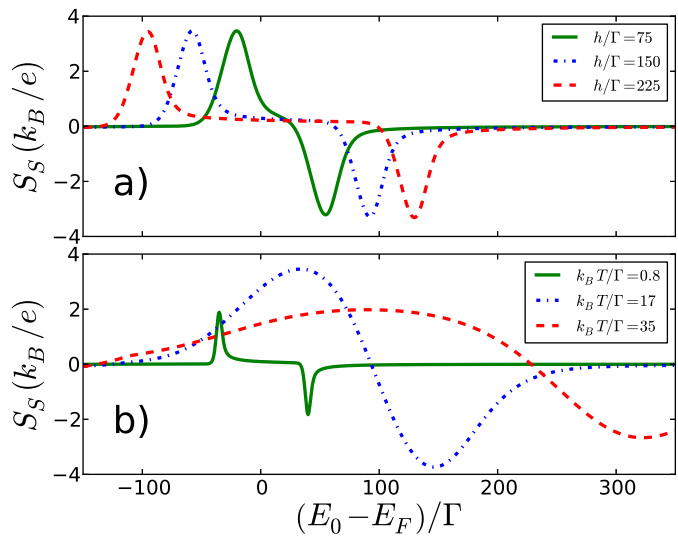}
 	\caption{(Color online) Spin Seebeck coefficient for a magnetic tunnel diode as a function of $E_0$ for (a) fixed temperature $k_B T/\Gamma = 4$ and different spin splittings $h$ and (b) fixed splitting $h/\Gamma=75$ and different temperatures.}
 	\label{fig: spin seebeck}
\end{figure}

In Fig. \ref{fig: spin seebeck} we plot the spin Seebeck coefficient for different magnetic fields (upper panel)
and different temperatures (lower panel). We can observe that unlike the charge thermopower, $S_S$ can be positive
or negative depending on the relative position of $E_0$ with respect to $E_F$.
Therefore, the spin bias due to a temperature difference ($V_S=-S_S \Delta T$) can change its sign
if we modify the position of $E_0$ using, e.g., nearby gates, doping or growing techniques.
This is not possible with charge degrees of freedom only (cf. Fig.~\ref{fig: seebeck}).
At low temperature, Fig.~\ref{fig: spin seebeck} shows that the spin thermopower is manifestly
positive (negative) for well level positions below (above) the Fermi energy.
The peak separation grows when $h$ increases, as expected.
In Fig.~\ref{fig: spin seebeck}(b) we analyze the influence of temperature for a fixed splitting $h$.
As the base temperature increases, the peak positions shift to higher energies but the maximum value
of $S_S$ stays roughly constant. In fact, for moderate temperatures ($T=35\Gamma/k_B$)
there is a wide range of energy levels around which the spin Seebeck coefficient
attains a sizeable value (around $0.2$~mV/K). Our numerical simulations
reveal a maximum of $|S_S|$ at $T\simeq 50K$ for the experimental values
$\Gamma=1$~meV, $E_F=50$~meV and $h=35$~meV.\cite{slo1}

To sum up, we have analyzed the thermoelectric properties of magnetic and nonmagnetic
resonant tunneling diodes using the scattering approach. We have found that in the absence
of magnetic fields the thermoelectric peak conductance occurs when the quantum well level
is aligned with the reservoirs' Fermi energy, and this effect persists when temperature changes.
In magnetically doped quantum wells, the spin thermopower can be tuned
with an external field and reaches significant values even if the background temperature increases.
Therefore, our work suggests that quantum well systems are quite promising in developing
substantial spin Seebeck effects at large output powers.
Electron-electron interactions will not qualitatively alter our conclusions since these interactions
are effectively screened in large-area heterostructures, although further work should take
into account the role of phonons and disorder in the intermediate temperature range.

This work has been supported by MINECO under grant No.~FIS2011-23526.

\end{document}